\documentclass[sigconf]{acmart}
\AtBeginDocument{%
  \providecommand\BibTeX{{%
    Bib\TeX}}
  \let\balance\relax}

\copyrightyear{2026}
\acmYear{2026}
\setcopyright{cc}
\setcctype{by}
\acmConference[Middleware '26]{27th International Middleware Conference}{December 14--18, 2026}{Tarragona, Spain}
\acmBooktitle{27th International Middleware Conference (Middleware '26), December 14--18, 2026, Tarragona, Spain}
\acmDOI{10.1145/3801927.3858633}
\acmISBN{979-8-4007-2621-7/2026/12}

\usepackage{algorithm}
\usepackage{algcompatible}
\usepackage[noend]{algpseudocode}

\usepackage{textcomp}
\usepackage{xspace}
\usepackage{multirow}
\usepackage{adjustbox}
\usepackage{booktabs}
\usepackage{array}
\usepackage{enumitem}
\usepackage{diagbox}
\usepackage[caption=false,font=footnotesize]{subfig}

\usepackage{needspace}
\usepackage{tikz}
\usetikzlibrary{arrows.meta,positioning,shapes.geometric}
\usepackage{pifont}
\usepackage{pgfplots}
\usepgfplotslibrary{statistics}
\usepackage{pgfplotstable}
\pgfplotsset{compat=1.18}

\usepackage{lipsum}
\usepackage{soul}
\usepackage{comment}
\usepackage{footnote}
\usepackage{siunitx}
\usepackage{etoolbox}
\usepackage{soulpos}
\usepackage{wrapfig}
\usepackage[utf8]{inputenc}
\usepackage{atbegshi}

\usepackage{mdframed}

\hypersetup{
    colorlinks=true,
    citecolor=blue,      
    linkcolor=blue,      
    urlcolor=blue,       
    breaklinks=true,
}

\AtBeginDocument{\microtypesetup{protrusion=true,expansion=true}}

\makesavenoteenv{tabular}
\makesavenoteenv{table}

\def\BibTeX{{\rm B\kern-.05em{\sc i\kern-.025em b}\kern-.08em
    T\kern-.1667em\lower.7ex\hbox{E}\kern-.125emX}}

\ulposdef{\gray}[xoffset=1pt]{\mbox{\color{gray!30}\rule[-.8ex]{\ulwidth}{3ex}}}
    
\newcommand\TAB{Table\xspace}
\newcommand\FIG{Figure\xspace}

\newcommand{\eg}{\textit{e.g.}}
\newcommand{\ie}{\textit{i.e.}}

\newcommand{\sys}{\textsf{FaaSpot}\xspace} 

\newcommand{\critical}{\texttt{CRITICAL}\xspace} 
\newcommand{\medium}{\texttt{MEDIUM}\xspace} 
\newcommand{\low}{\texttt{LOW}\xspace} 

\newcommand{\wfcnt}{{eight}\xspace} 

\def\footnoterule{\relax%
  \kern0pt
  \hbox to \columnwidth{\hfill\vrule width 1.0\columnwidth height 0.4pt\hfill}
  \kern4.6pt}

\newcommand{\para}[1]{\vspace{1mm}\noindent{\bf #1.}}

\definecolor{fgray}{gray}{0.9}
\definecolor{auburn}{rgb}{0.43, 0.21, 0.1}
\definecolor{rblue}{rgb}{0.84, 1.0, 1.0}
\definecolor{ao}{rgb}{0.0, 0.0, 1.0}
\definecolor{navyblue}{rgb}{0.0, 0.0, 0.5}

\newcounter{mlpcounter}
\newcounter{ikkcounter}
\newcounter{asmcounter}
\newcounter{tcounter}
\newcounter{Xcounter}  
\newcounter{Rcounter}

\newmdenv[
  backgroundcolor=yellow!20,
  linecolor=black,
  linewidth=1pt,
  roundcorner=5pt
]{takeaway}

\begin{document}

\title{\sys: Fault-Tolerant Serverless Workflow Execution on Spot Instances}
\author{Bandhan Patel}
\affiliation{%
  \institution{University of Georgia}
  \city{Athens, GA}
  \country{USA}
}
\email{bandhan.patel@uga.edu}

\author{Sushruth Harsha}
\affiliation{%
  \institution{University of Georgia}
  \city{Athens, GA}
  \country{USA}
}
\email{sh90804@uga.edu}

\author{Kwangsung Oh}
\affiliation{%
  \institution{University of Nebraska at Omaha}
  \city{Omaha, NE}
  \country{USA}
}
\email{kwangsungoh@unomaha.edu}

\author{In Kee Kim}
\affiliation{%
  \institution{University of Georgia}
  \city{Athens, GA}
  \country{USA}
}
\email{inkee.kim@uga.edu}

\begin{abstract}
Serverless computing is increasingly used to run complex DAG workflows across diverse applications, and individual function execution times have grown to support them, pushing workflow makespans to hours. 
Open-source serverless platforms, deployed on a pool of VMs, typically use on-demand instances, and their cost increases with the makespan.
Spot instances are far cheaper, but a preemption is likely before the workflow finishes. 
{A preemption at a fan-out node or a task feeding a fan-in node stalls every parallel branch, cascading a single preemption into a workflow-wide failure.}
Prior systems address this challenge by migrating high-impact tasks to reliable, non-spot resources, sacrificing spot's cost advantage. 

We present \sys, a serverless workflow execution system that 
runs every workflow task on spot instances with no on-demand fallback.
Our key insight is that cascade-prone positions are determined by the DAG's topology and can be identified before execution. 
Therefore, \sys analyzes the workflow DAG once at submission time and classifies each task by the impact of its preemption. 
Selective checkpointing, survival-aware scaling, and lineage-based retry then act on this single classification, with no runtime coordination. 
We evaluate \sys on a trace-driven spot emulator with eight DAG workflows against four state-of-the-art baselines.
\sys completes 89.03\% of runs at \$0.018 per successful run, 
21pp (percentage points) above the average baseline success rate and at a lower cost per success than every baseline, 
while reducing mean latency by 2.3$\times$.
\end{abstract}


\maketitle

\section{Introduction}\label{sec:intro}

Serverless computing is increasingly used to execute complex directed acyclic graph (DAG) workflows in data analytics \cite{dexter.middleware24, jolton.nsdi24, minflow.fast24}, machine learning \cite{smiless.sc24, mbs.vldb22, cirrus.socc19}, and scientific computing \cite{wukong.socc20, daydream.sc22, mashup.ppopp22}. 
These workflows are submitted, run to completion, and terminate, and they typically run repeatedly on a schedule or per experiment.
To support such applications, major cloud providers now offer dedicated orchestration services \cite{stepfunctions.web26, durablefunctions.web26, googleworkflow.web26}, 
and individual function execution times have grown substantially. 
For example, Google Cloud now allows a single function to run for up to 60 minutes \cite{gcpfuncduration.web26}, while Azure has removed execution time limits for its current plans \cite{azurefuncduration.web26}, compared to the few-minute limits of earlier generations.
As a result, longer function chains and extended per-function runtimes push workflow makespans from tens of minutes to several hours.

\begin{figure}[t]
  \centering
  \includegraphics[width=0.55\linewidth]{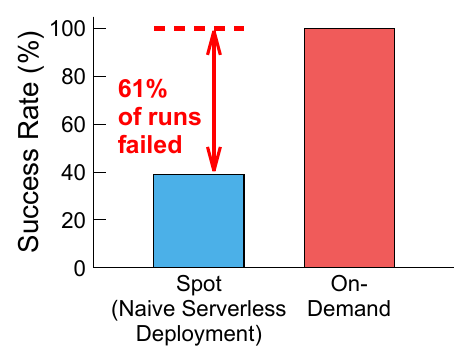}
  \vspace{-0.5em}
  \caption{Average success rate of naive serverless deployment on spot vs. on-demand.  
  {\normalfont \em Results are averaged across our eight-workflows (500 runs each). Per-workflow results appear in \FIG\ref{fig:spot-reliability-v1}.}}
  \label{fig:avg-failure}
  \vspace{-0.5em}
\end{figure}

Alongside these managed offerings, self-managed open-source serverless platforms, \eg, Fission \cite{fission.web26}, OpenWhisk \cite{openwhisk-composer.web26}, are increasingly adopted, where operators run the serverless infrastructure on their own pool of VMs. 
These VMs are typically provisioned as on-demand instances, so 
infrastructure costs increase with workflow duration. 
Spot instances (2$\times$ -- 10$\times$ cheaper \cite{cantbelate.nsdi24, spotmarketnotsustain.hotcloud16, pricenotright.icccn19}) become an attractive option for significantly reducing this cost. 
However, for workflows that run for tens of minutes to hours, spot instances are likely to be preempted before they finish \cite{cantbelate.nsdi24, lee.iiswc22}.

\begin{figure*}[t]
  \centering
  \subfloat[Success rate across workflows]{%
    \includegraphics[width=0.32\linewidth]{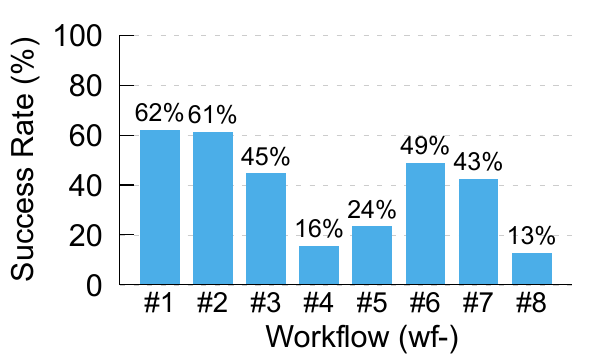}%
    \label{fig:wf-success}}%
  \hfill
  \subfloat[Cost per successful run]{%
    \includegraphics[width=0.32\linewidth]{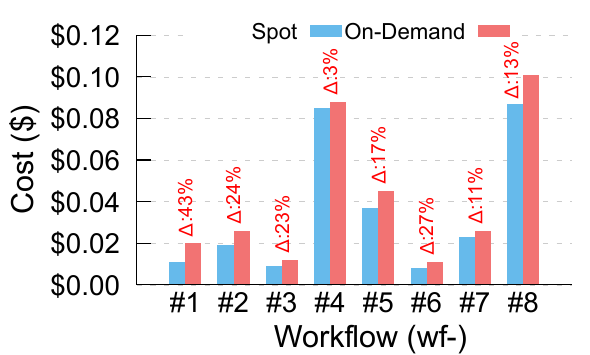}%
    \label{fig:wf-cost}}%
  \hfill
  \subfloat[Average end-to-end latency]{%
    \includegraphics[width=0.32\linewidth]{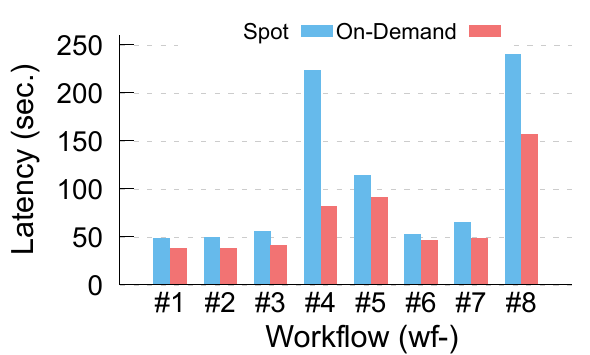}%
    \label{fig:wf-latency}}%
    \vspace{-0.5em}
  \caption{Success rate, cost, and latency of naive serverless deployment on spot, compared to on-demand, across our \wfcnt-workflow suite.
  {\normalfont \em
  500 runs per workflow on a trace-driven spot emulator (\S\ref{subsec:spot-emul}), using SkyPilot lifetime traces \cite{cantbelate.nsdi24} for {\tt p3.2xlarge} spot instances and SpotLake \cite{lee.iiswc22} pricing.
  (a) `Success rate' falls significantly below 50\% on the more complex workflows, dropping as low as $\sim$13\%, while on-demand completes every run. 
  {(b) `Cost-per-successful-run' stays below on-demand on every workflow, but the margin ($\Delta$) is modest. Apart from the simplest workflow (43\%), it stays below 30\% and falls as low as 3\%, far below spot's nominal price discount.}
  (c) `End-to-end latency on spot' exceeds on-demand by up to 2.7$\times$ because every failure forces a restart from the workflow's origin (the first function of the DAG).}}
  \label{fig:spot-reliability-v1}
  \vspace{-0.5em}
\end{figure*}

We measured the impact of spot preemptions on workflow completion. 
We deployed a set of serverless workflows naively on spot VMs, without any fault-tolerance mechanism.
As shown in \FIG\ref{fig:avg-failure}, we observed that they completed only 39\% of their runs on average, whereas on-demand deployments completed every run. 
The impact of a preemption is highly dependent on where it occurs in the workflow. 
For example, a preemption on a sequential {\em task}\footnote{We use `task' and `function' interchangeably, since each task in a serverless workflow runs as a function.} (or {\em function}) affects only that task, whereas a preemption at a fan-out node or a task feeding a fan-in node can stall every parallel branch that shares it, a {\em cascade} that turns a single failure into many.
As a result, a small number of structurally important positions are responsible for most of the workflow failures.

Fault tolerance on spot and transient resources has been studied mostly for non-serverless workloads \cite{spotverse.middleware24, flint.eurosys16, sharma.sigmetrics17, tributary.atc18}.
Most existing systems do not consider where preemptions occur in the workflow. 
They protect every task uniformly or react to preemptions at the VM level, without distinguishing a structurally critical task from an ordinary one (\eg, a task on a sequential path). 
A few systems (\eg, TR-Spark \cite{trspark.socc16}, Pado \cite{pado.eurosys17}) analyze the workload's DAG to protect the tasks whose preemption is most costly.
But these systems all rely on reliable resources without preemption risk, migrating these high-impact tasks, those that can trigger a cascade, to stable storage or reserved nodes. 
Such use of non-spot resources (even if partial) reduces the spot's cost advantage. 
To fully leverage that advantage, as much of the workload as possible (ideally all of it) should run on spot. 
This direction leaves no reliable place to host or migrate these tasks. 
Therefore, the structural insight from above must be fully implemented for serverless workflows on spot, without the reliable resources that prior approaches assume.

To address this challenge, we present \sys, a serverless workflow execution system that 
runs every workflow task on spot instances, with no on-demand fallback.
\sys's key insight is that positions (where a preemption would trigger a cascade) can be identified before execution begins, as they are determined by the DAG's topology rather than runtime state. 
\sys analyzes the DAG once at submission time and classifies each task by the impact of its preemption. 
This single offline classification drives three online mechanisms: 
1) selective checkpointing persists state only at high-impact positions; 
2) survival-aware scaling provisions safer instances before a preemption is likely; 
and 3) lineage-based retry resumes a failed run from its nearest checkpoint rather than its origin. 
Because all three mechanisms are guided by the same classification, they consistently protect high-impact tasks, with no runtime coordination and no migration to non-spot resources. 
The result is spot-level cost with the reliability of on-demand execution.

We implement \sys as a working prototype and develop a trace-driven spot emulator to evaluate it. 
Our workflow suite spans various application domains and a range of structures, from simple fork-join graphs to complex real-world pipelines~\cite{workflows.works08, mashup.ppopp22, 1000genomes.nature, gridgreen.socc25}, including Epigenomics, LIGO, and Montage. 
Ensuring that \sys and competing baseline systems receive identical spot preemption events is extremely challenging on live spot instances, so the emulator replays real spot lifetime and pricing traces at high fidelity.
The emulator is also decoupled from \sys, enabling any spot-aware system to be evaluated under the same preemption events. 
Across \wfcnt workflows and against four state-of-the-art spot fault-tolerance systems \cite{snape.asplos23, hourglass.eurosys19, mscheduler.cloudcom23, teylo.tcc23}, 
\sys completes 89.03\% of runs on average at \$0.018 per successful run, 21pp (percentage points) above the average baseline success rate and at a lower cost per success than every baseline. 
Even against the strongest baseline (81.78\% at \$0.040), \sys completes more runs at less than half the cost while reducing mean latency by 2.3$\times$, outperforming all four on every evaluation metric.

This work makes the following contributions toward fault-tolerant execution of serverless workflows on spot instances.
\begin{itemize}[leftmargin=1em]
  \item {\bf \sys}, a serverless workflow execution system that 
  executes every serverless workflow task on spot instances without on-demand fallback (\S\ref{sec:system}).
  \item {\bf A unified failure-aware execution framework} that uses a single offline task classification to drive selective checkpointing, survival-aware scaling, and lineage-based retry, enabling consistent protection of high-impact tasks without runtime coordination (\S\ref{sec:system_components}).
  \item {\bf A trace-driven evaluation harness} that replays real spot lifetime and pricing traces, enabling fair and reproducible comparisons among spot-aware workflow execution systems under identical preemption events (\S\ref{subsec:spot-emul}, \S\ref{sec:eval-v1}).
\end{itemize}

\section{Motivation}\label{sec:motivation}

\begin{figure*}[t]
  \centering
  \includegraphics[width=0.9\linewidth]{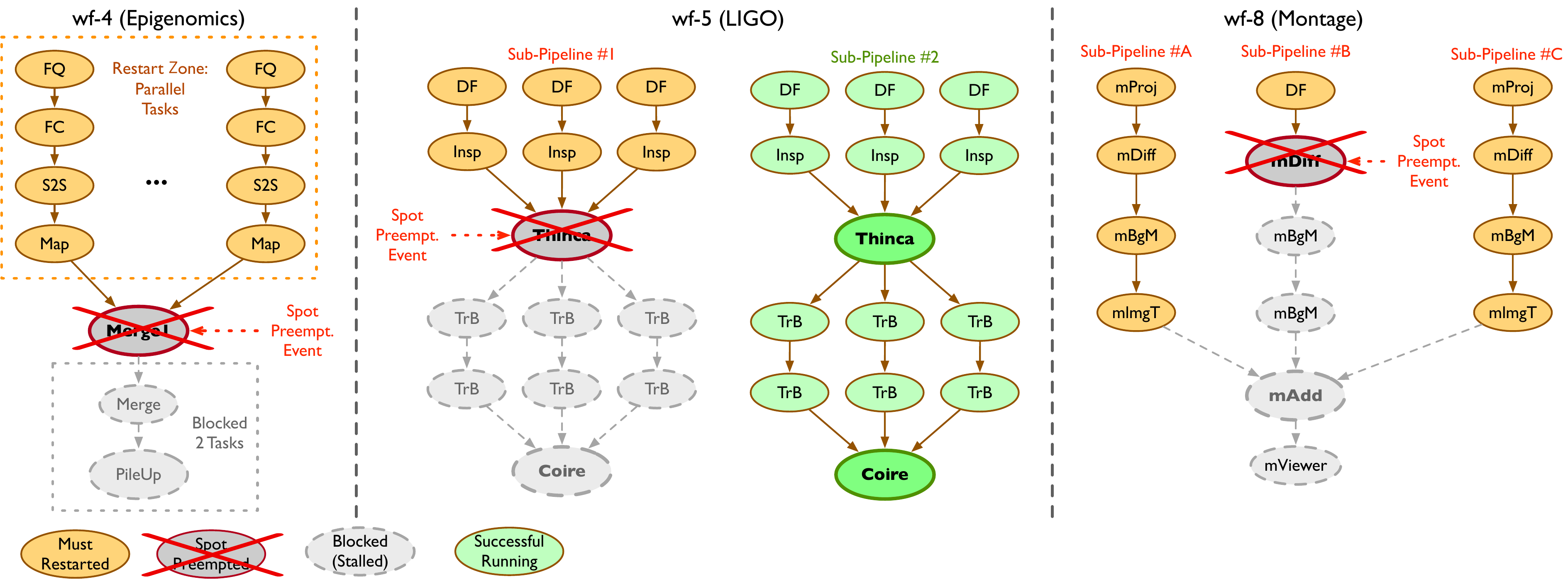}
  \vspace{-0.5em}
  \caption{Three representative DAG workflows and the cascade failures at their synchronization barriers. 
  {\normalfont \em wf-4 (Epigenomics~\cite{workflows.works08}), wf-5 (LIGO~\cite{workflows.works08}), and wf-8 (Montage~\cite{workflows.works08}) each contain fan-out nodes and synchronization barriers. Under naive deployment, a preemption of any parallel task feeding a barrier fails the entire run: the preempted task (red), every other branch that must restart with it (orange), and the downstream tasks left stalled at the barrier (gray). For wf-5, the two sub-pipelines contrast a preemption-hit branch with an unaffected one.}}
  \label{fig:three-wfs}  
  \vspace{-0.5em}
\end{figure*}

\subsection{Serverless Workflow Deployment on Spot}\label{subsec:wfonspot}

Spot instances are significantly cheaper than on-demand \cite{cantbelate.nsdi24, spotmarketnotsustain.hotcloud16, pricenotright.icccn19}, which makes them an attractive model for serverless DAG workflows. 
To quantify the actual cost benefits and reliability risk, we deployed a set of DAG workflows directly on spot instances using the serverless platform's default configuration, with no checkpointing, no retry coordination, and no migration to on-demand. 
We refer to this configuration as naive serverless deployment. A preemption anywhere in the workflow fails the entire run.

The measurements use a trace-driven spot emulator (described in \S\ref{subsec:spot-emul}). 
The emulator replays SkyPilot lifetime traces \cite{cantbelate.nsdi24} for {\tt p3.2xlarge} spot instances. 
Cost is computed using SpotLake \cite{lee.iiswc22} live spot pricing. Each workflow is run 500 times under matched trace conditions, allowing direct paired comparison against on-demand execution of the same workflow.

\FIG\ref{fig:spot-reliability-v1} reports three measurements comparing naive spot deployment against on-demand across our \wfcnt-workflow suite\footnote{Three workflows are detailed in \FIG\ref{fig:three-wfs}, with the full suite described in \S\ref{sec:eval-v1}.}.
Success rate of the naive deployment (\FIG\ref{fig:wf-success}) falls well below 50\% on the more complex workflows, dropping as low as $\sim$13\%, while on-demand completes every run. 
Cost-per-successful-run (\FIG\ref{fig:wf-cost}) stays below on-demand on every workflow (but only by a modest margin).
End-to-end latency (\FIG\ref{fig:wf-latency}) exceeds on-demand by up to 2.7$\times$ because every failure forces a restart from the workflow's origin (the first function).

The three measurements show spot VM's strengths and weaknesses. 
A cost-benefit remains, but it is modest. 
Spot is cheaper per successful run on every workflow, but except for the simplest one, the gap stays below 30\% and falls as low as 3\% (\FIG\ref{fig:wf-cost}), far below the 2$\times$ to 10$\times$ discount reported in prior work \cite{cantbelate.nsdi24, spotmarketnotsustain.hotcloud16, pricenotright.icccn19}. 
Repeated restarts of failed workflows due to spot preemption diminish spot's price advantage. 
Furthermore, the increased latency and high failure rate are direct and substantial. 
This gap clearly indicates our opportunity. 
A fault-tolerant mechanism could recover spot's full price advantage while significantly reducing latency and failure rates to on-demand levels.


\subsection{Why Complex Workflows Fail More Often?}\label{subsec:complexwf}

The high failure rate on complex workflows (\FIG\ref{fig:wf-success}) cannot be explained by task count or execution time alone. 
The primary cause is dependency structure, specifically fan-out nodes and synchronization barriers. 
A preemption at a fan-out node blocks every parallel branch it spawns. 
A synchronization barrier is where all parallel branches must complete and merge at a fan-in node before downstream execution can continue, so a preemption of any task feeding the barrier stalls every other branch waiting at it. 
Three workflows exhibit both patterns, 
wf-4 (Epigenomics~\cite{workflows.works08}), 
wf-5 (LIGO~\cite{workflows.works08}), 
and wf-8 (Montage~\cite{workflows.works08}), shown in \FIG\ref{fig:three-wfs}.

\begin{figure}[t]
  \centering
  \includegraphics[width=0.8\linewidth]{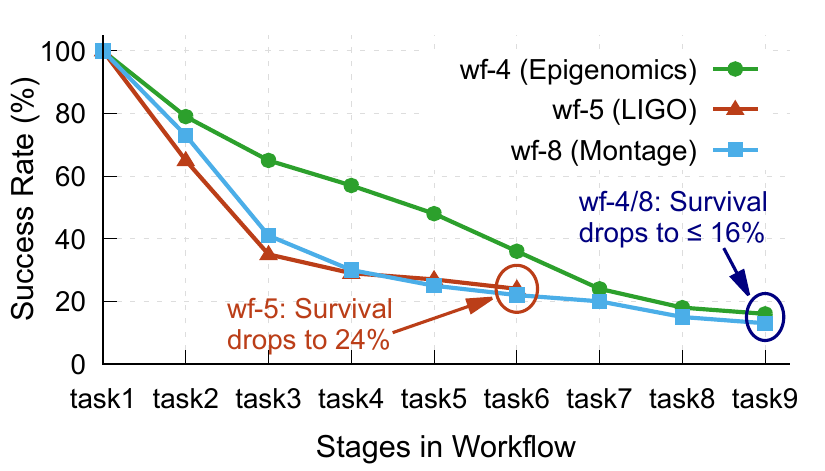}
  \vspace{-0.5em}
  \caption{Stage-level success rate for three complex workflows under naive serverless deployment on spot.
  {\normalfont \em All three workflows start near 100\% at the first stage and decrease monotonically, with 13\%--24\% of runs surviving to the final stage.}}
  \label{fig:success-by-stage}
  \vspace{-1em}
\end{figure}

\FIG\ref{fig:success-by-stage} decomposes each workflow's success rate by stage, measured over 500 trials of naive deployment. 
All three workflows start near 100\% at the first stage and decrease monotonically. 
However, only 13\% to 24\% of workflow runs survive at the final stage.
The steepest drops occur in the early stages, where the full population of runs is exposed to preemption across active parallel branches. 
wf-5 falls from 100\% to 37\% within three stages, and wf-8 falls to 42\% over the same span.
Each additional stage provides another opportunity for preemption to occur somewhere in the workflow, and any preemption can cause the entire run to fail.
The cost is not uniform across DAG positions. 
A preemption on a sequential path forces only that task to recompute, but a preemption at a fan-out node or at a synchronization barrier propagates to every branch that shares it.

Complex workflows fail more often because they contain more high-impact positions. They have more fan-out nodes, wider parallel fan, and deeper synchronization barriers.
Therefore, the same level of preemption activity produces a higher cumulative failure cost in complex workflows than in simple ones.

The pattern is not specific to any single application domain. 
wf-4 (Epigenomics), wf-5 (LIGO), and wf-8 (Montage) have distinct DAG topologies, yet all three suffer similar cascade failure rates. 
Because high-impact preemption positions are statically derivable from the DAG before execution begins, they can be selectively protected rather than uniformly.

\begin{figure}[t]
  \centering
  \includegraphics[width=0.85\linewidth]{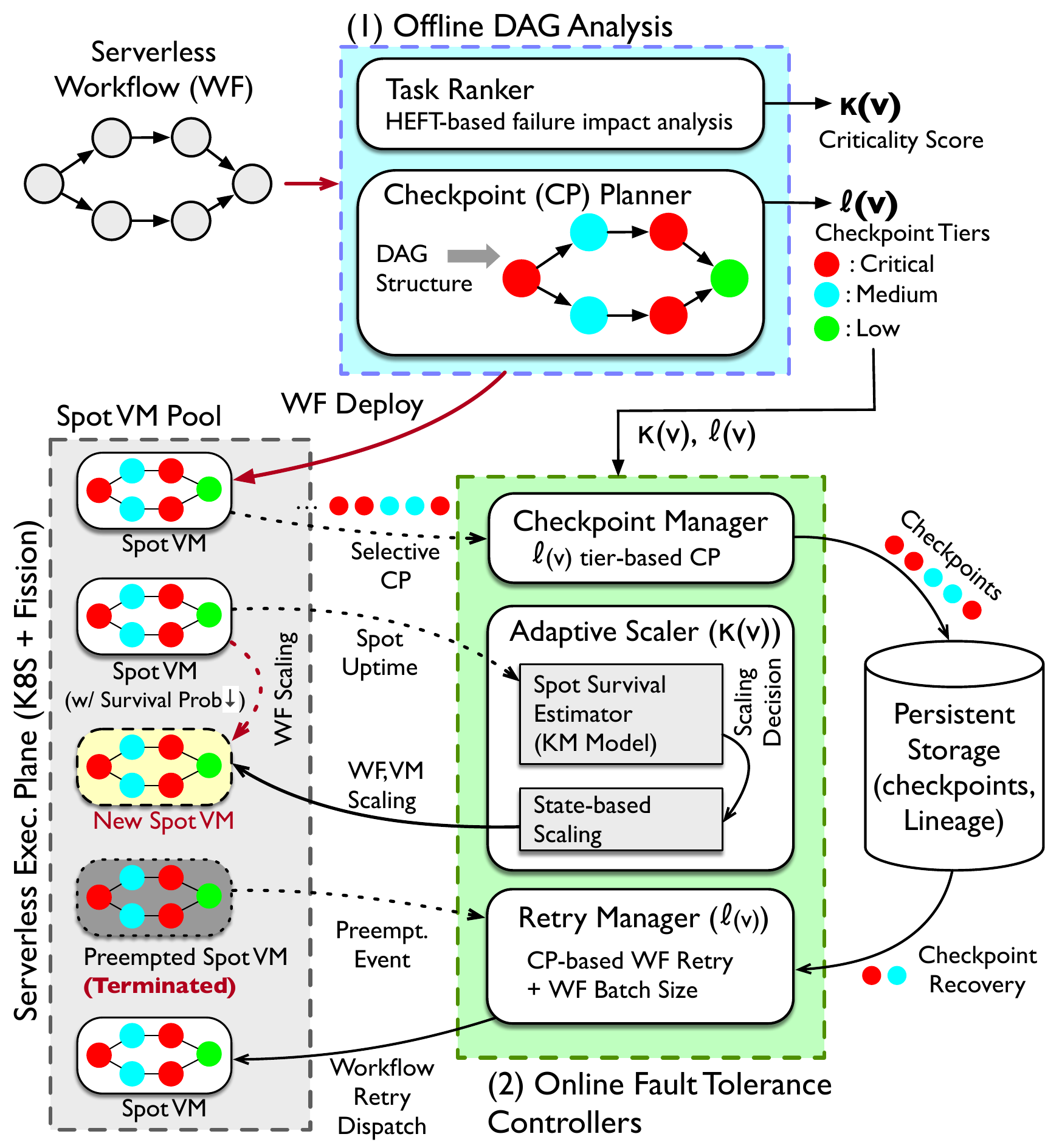}
  \vspace{-0.5em}
  \caption{\sys system architecture.
  {\normalfont \em 
  Offline phase (blue, (1)) derives a criticality score $\kappa(v)$ and checkpoint tier $\ell(v)$ from the workflow DAG. 
  Three online fault-tolerance controllers (green, (2)) consume these signals at runtime (\S\ref{subsec:chkptmgr}--\S\ref{subsec:retry}). 
  Execution plane (left) runs task pods on a shared spot pool via Fission on Kubernetes.}}
  \label{fig:checkscale-arch}
  \vspace{-1em}
\end{figure}

\section{\sys System Overview}\label{sec:system}

\sys enables complex serverless workflows to run on spot instances by tolerating the cascade failures characterized in \S\ref{sec:motivation}. 
\sys achieves spot-level cost with on-demand-level reliability, without migrating any task off spot. 
This section overviews its architecture (\S\ref{subsec:arch}) and the three design principles behind it (\S\ref{subsec:design}): 
{\em offline DAG analysis} to identify and classify high-impact preemption positions, 
{\em a criticality-annotated DAG} that encodes per-task structural risk and is shared across all fault tolerance decisions, and 
{\em tiered fault tolerance} that matches protection intensity to each task's cascade potential.


\subsection{\sys Architecture}\label{subsec:arch}

\FIG\ref{fig:checkscale-arch} shows the architecture of \sys, which takes a serverless workflow DAG $G=(V,E)$ as input and executes it on spot instances through two (offline and online) coordinated phases.

The {\bf offline phase} consists of two components. 
The {\em task ranker} (\S\ref{subsec:dag-analysis}) computes a per-task criticality score $\kappa(v)$ that quantifies each task's downstream failure impact. 
The {\em checkpoint planner} (\S\ref{subsec:dag-analysis}) classifies each task into three tiers $\ell(v)$ (\low, \medium, \critical in \TAB\ref{tab:criticality}) based on its structural position in the DAG.
The criticality score $\kappa(v)$ and tier $\ell(v)$ are computed once before execution begins, then passed to all online controllers.

The {\bf online phase} manages fault tolerance during execution through three controllers: the {\em checkpoint manager} (\S\ref{subsec:chkptmgr}), the {\em adaptive scaler} (\S\ref{subsec:scaler}), and the {\em retry manager} (\S\ref{subsec:retry}).
All three make decisions based on the same $\kappa(v)$ and $\ell(v)$ from the offline phase.

\sys runs on top of a serverless execution plane and persistent storage. The execution plane uses Fission \cite{fission.web26}, an open-source serverless platform, as the control plane and runs each task as a Kubernetes \cite{kubernetes.web26} pod on a shared spot VM pool (\FIG\ref{fig:checkscale-arch}, left). 
The persistent storage holds checkpoints and lineage records used by the online fault-tolerance controllers for recovery.

\para{{Scope and assumptions}} 
{\sys runs every workflow task on spot with no on-demand fallback. 
However, the persistent storage and the Fission control plane are not on spot, and neither grows with the workflow. 
The design assumes that the DAG is fixed at submission and that a per-task execution time estimate is available.}

\para{\sys execution flow}
As the execution plane dispatches tasks, 
the checkpoint manager checks each task's tier $\ell(v)$. 
Checkpoints are written for tasks in higher tiers and skipped for those in the lowest tier to reduce overhead.
Concurrently, the adaptive scaler monitors each spot instance's survival probability based on its uptime. 
When preemption risk increases, it scales out workflow tasks onto safer instances, provisioning new VMs if the safer-instance pool nears capacity.
When a spot instance is preempted, the retry manager identifies the affected tasks and resumes each from its latest checkpoint rather than restarting from the workflow's origin. 
The adaptive scaler may concurrently provision additional resources to maintain forward progress. 
Execution continues until all workflow tasks complete.

\subsection{Three Design Principles}\label{subsec:design}

We further elaborate on the three principles of \sys, explaining why each design choice is made and how these choices reinforce one another.

\para{Offline DAG analysis}
Runtime identification of high-impact preemption positions often requires continuous analysis of the DAG as tasks execute, which can add latency to every scheduling decision. 
\sys avoids this overhead by relying on a key property: fan-in and fan-out nodes are topological features that do not depend on runtime state. They can be identified once, before execution begins, and the result reused across all fault tolerance decisions.

\para{Criticality-annotated DAG}
The offline DAG analysis annotates each task with its criticality score $\kappa(v)$ and tier $\ell(v)$, producing a {\em criticality-annotated DAG}. 
This annotation is the only state the three online controllers share. 
Because every fault-tolerance decision is based on this same offline classification rather than on independently maintained runtime state, their decisions are consistent without any runtime coordination. 
\S\ref{sec:system_components} details how each controller consumes $\kappa(v)$ and $\ell(v)$.

\begin{table}[t]
  \centering
  \small
  \caption{Tiered fault tolerance policy by task classification.
  {\normalfont \em Each task's $\ell(v)$ (tier) determines whether its state is checkpointed and therefore whether a preempted run can resume from it.}}
  \vspace{-0.5em}
  \resizebox{1\columnwidth}{!}
  {
	\begin{tabular}{llll}
	\toprule
	\multicolumn{1}{l}{\textbf{Tier}} &
	  \multicolumn{1}{l}{\textbf{\begin{tabular}[c]{@{}c@{}}Structural Location\end{tabular}}} &
	  \multicolumn{1}{l}{\textbf{\begin{tabular}[c]{@{}c@{}}Ckpt.\end{tabular}}} &
	  \multicolumn{1}{l}{\textbf{\begin{tabular}[c]{@{}c@{}}{Recovery Point}\end{tabular}}} \\ \midrule
	{\critical} &
	  \begin{tabular}[c]{@{}l@{}}Fan-out node, or task\\ feeding a fan-in node\end{tabular} &
	  \ding{51} &
	  \begin{tabular}[c]{@{}l@{}} {Yes, preserves all}  \\ {branches past the barrier} \end{tabular} \\ \midrule
	{\medium} &
	  \begin{tabular}[c]{@{}l@{}}Parallel task that is neither \\ a fan-out node nor feeds \\a fan-in node\end{tabular} &
	  \ding{51} &
	  \begin{tabular}[c]{@{}l@{}} {Yes, preserves its} \\ {own branch} \end{tabular}\\ \midrule
	{\low} &
	  \begin{tabular}[c]{@{}l@{}}Sequential path, or fan-in \\ node itself \end{tabular} &
	  \ding{55} &
	  \begin{tabular}[c]{@{}l@{}} {No, replayed during} \\ {recovery} \end{tabular} \\
	  \bottomrule
	\end{tabular}
  }
  \label{tab:criticality}
  \vspace{-0.5em}
\end{table}

\para{Tiered fault tolerance}
A uniform checkpoint policy treats all tasks equally, incurring I/O overhead on sequential tasks where re-execution cost is bounded by a single task. 
\sys instead classifies tasks into three tiers based on the scope of cascade their preemption can trigger (\TAB\ref{tab:criticality}).
This classification allows the checkpoint manager to focus checkpoint I/O on high-impact locations and the retry manager to prioritize recovery for tasks with the highest downstream impact.

\section{Failure-aware Serverless Workflow Execution}\label{sec:system_components}

Building on the three design principles introduced in \S\ref{subsec:design}, we describe \sys's fault-tolerant execution mechanisms for serverless workflows on spot instances. 
\S\ref{subsec:dag-analysis} describes the offline DAG analysis that produces the criticality score $\kappa(v)$ and checkpoint tier $\ell(v)$ before execution begins. 
The remaining three subsections describe the online fault-tolerance controllers: 
the checkpoint manager (\S\ref{subsec:chkptmgr}), which writes checkpoints at tier-selected positions, 
the adaptive scaler (\S\ref{subsec:scaler}), which provisions safer VMs when survival probability drops, and 
the retry manager (\S\ref{subsec:retry}), which resumes preempted workflows from their nearest persisted checkpoint. 
All components operate on the same criticality-annotated DAG produced by the offline analysis.

\subsection{Offline DAG Analysis}\label{subsec:dag-analysis}
Before execution begins, \sys analyzes the workflow DAG using two components: the {\em task ranker} and the {\em checkpoint planner}.
The task ranker computes a criticality score $\kappa(v)$, and 
checkpoint planner assigns a checkpoint tier $\ell(v) \in \{\text{\critical}, \text{\medium}, \text{\low}\}$ (Algo. \ref{alg:checkpoint-classifier}). 
Both $\kappa(v)$ and $\ell(v)$ are computed once and passed to the online controllers (\FIG\ref{fig:checkscale-arch}).

\para{Task ranker}
The criticality score $\kappa(v)$ reflects both the longest remaining path from $v$ and its structural position. 
A high $\kappa(v)$ means a preemption at $v$ stalls many subsequent tasks, while a low $\kappa(v)$ means it affects only a few. 
\sys builds on the Heterogeneous Earliest Finish Time (HEFT) algorithm, a standard priority metric in DAG scheduling that captures the longest remaining path from each task \cite{heft.tpds02}. 
The base HEFT rank is formulated as:
\begin{equation}
\mathrm{rank}(v)=w(v)+\max_{u\in\mathrm{succ}(v)}\mathrm{rank}(u),
\label{eq:rank}
\end{equation}
where $w(v)$ is the execution time of $v$. 
The task ranker then adds two structural counts to $rank(v)$, 
the number of branches $v$ creates and the number stalling at the barrier $v$ feeds, 
and min-max normalizes the sum across all tasks to obtain $\kappa(v) \in [0, 1]$. 
A fan-out node already carries a high $rank(v)$ via the longest downstream chain, and the degree terms separate tasks whose remaining paths are comparable. 
The adaptive scaler (\S\ref{subsec:scaler}) uses only the resulting order to prioritize critical tasks during scale-out.


\begin{algorithm}[t]
\caption{Checkpoint Classification}
\label{alg:checkpoint-classifier}
\begin{algorithmic}[1]
\Require Tasks $V$, with $\text{succ}(v)$ and $\text{minScale}(v)$ for each $v \in V$
\Ensure Checkpoint levels $\ell(v) \in \{\text{\critical}, \text{\medium}, \text{\low}\}$
\ForAll{$v \in V$}
    \If{$\exists\, s \in \text{succ}(v) : \text{minScale}(s) \neq \text{minScale}(v)$}
        \State $\ell(v) \gets \text{\critical}$
    \ElsIf{$\text{minScale}(v) > 1$}
        \State $\ell(v) \gets \text{\medium}$
    \Else
        \State $\ell(v) \gets \text{\low}$
    \EndIf
\EndFor
\State \Return $\ell$
\end{algorithmic}
\end{algorithm}

\begin{figure}[h]
  \centering
  \includegraphics[width=0.95\linewidth]{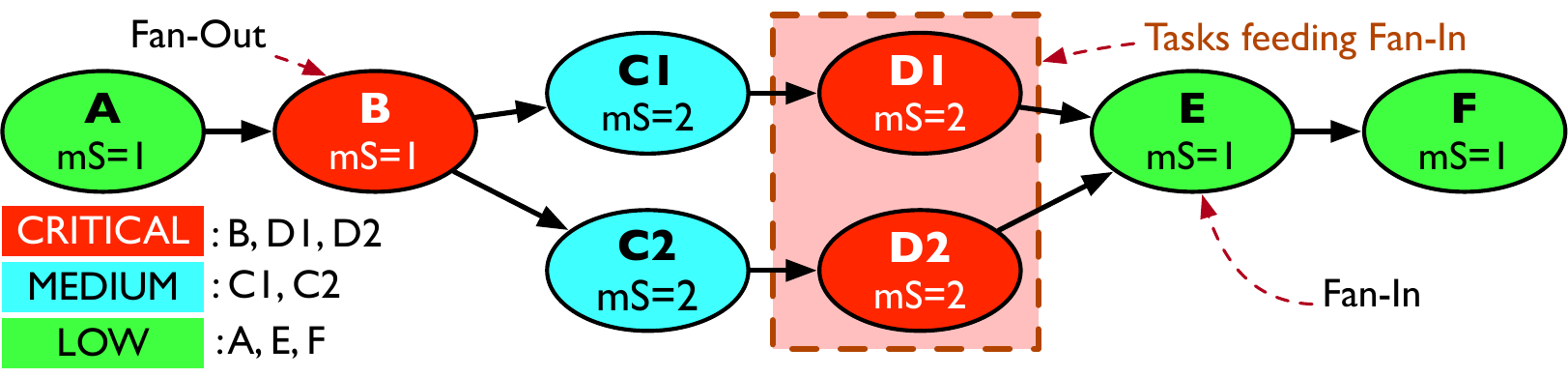}%
  \vspace{-0.5em}
  \caption{Three-tier $\ell(v)$ classification on an example DAG
  {\normalfont \em 
  \critical: fan-out node (B) or tasks feeding a fan-in barrier (D1, D2, shaded). 
  \medium: parallel-branch interior (C1, C2). 
  \low: sequential or fan-in nodes (A, E, F).}}
  \label{fig:dag-example}
  \vspace{-0.5em}
\end{figure}

\para{Checkpoint planner}
This component assigns each task $v$ a tier $\ell(v)$ (\critical, \medium, \low) in the workflow DAG.
Algo. \ref{alg:checkpoint-classifier} formalizes this classification using $v$'s minScale. 
The minScale of $v$ is the number of parallel tasks (functions) $v$ requires.
\ie, 1 for a sequential task, or $N$ for a task that runs as $N$ parallel branches. 
\critical tasks are fan-out nodes or parallel tasks feeding a fan-in node ($v$'s successor has a different minScale).
\medium tasks have similar parallelism as their successors (minScale $>$ 1, same as successors). 
\low tasks lie on sequential paths or are fan-in nodes themselves (minScale = 1, same as successors).
{A change in minScale marks exactly the positions we identified as cascade-prone (\S\ref{subsec:complexwf}). 
Therefore, the tier signal captures where a preemption propagates across branches rather than where re-execution is merely expensive.} 
\FIG\ref{fig:dag-example} illustrates the classification on an example DAG.
The checkpoint planner computes $\ell(v)$ once offline, in parallel with the task ranker.

\subsection{Selective Checkpointing}\label{subsec:chkptmgr}
At runtime, when a task (a node/function in a single workflow) completes, the checkpoint manager checks its assigned tier $\ell(v)$ and decides whether to write a checkpoint to persistent storage. 
\critical and \medium tasks receive a checkpoint upon completion, while LOW tasks are skipped (\TAB\ref{tab:criticality}). 
This selective checkpointing concentrates checkpoint I/O at cascade-prone positions in the DAG, where a persisted checkpoint recovers the most downstream work, and avoids the overhead of writing checkpoints at sequential tasks where re-execution cost is bounded by a single task.

\subsection{Survival-aware Scaling}\label{subsec:scaler}

\sys's survival-aware scaling is conducted by the adaptive scaler, which monitors the survival probability of spot instances and provisions additional spot VMs when the preemption risk increases. 
The adaptive scaler component combines three mechanisms:
a Kaplan-Meier model estimates per-instance survival probability from observed uptime, 
a state-based policy determines the scaling factor,
and HEFT ranks (computed offline by the task ranker, \S\ref{subsec:dag-analysis}) determine the scale-out sequence, prioritizing critical tasks on at-risk spot instances.

\begin{figure}[t]
  \centering
  \includegraphics[width=0.8\linewidth]{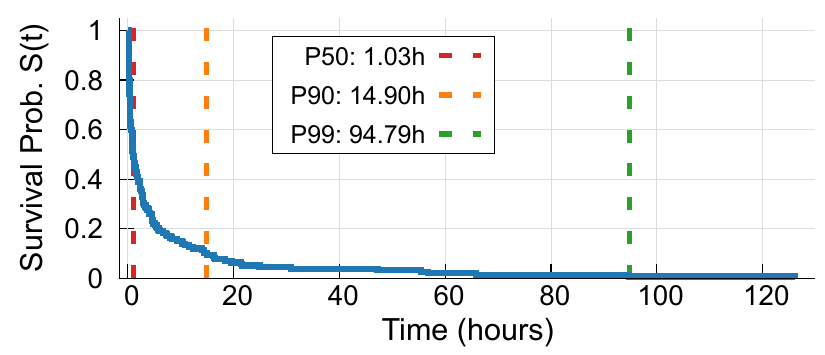}
  \vspace{-0.5em}
  \caption{Conditional survival curve for p3.2xlarge spot instances in AWS region {us-west-2a}. 
  {\normalfont \em The survival probability $S(t)$ falls steeply within the first few hours, then tails off. \sys's adaptive scaler maps each instance's $S(t)$ to a survival state (Table \ref{tab:survival-states}) and triggers scale-out before spot preemption.}}
  \label{fig:spot-survival}
  \vspace{-0.5em}
\end{figure}

\begin{table}[t]
  \centering
  \small
  \caption{Survival states (0--3) and corresponding scaling factor ranges in the adaptive scaler.
  {\normalfont \em Spot instances in the `Safe (0) state' do not require scale-out (Algo. \ref{alg:scaler}, lines 7--8); the other three states map $S(t)$ to a factor within the listed range.}}
  \label{tab:survival-states}
  \vspace{-0.5em}
  {%
  \begin{tabular}{lll}
  \toprule
  \textbf{Survival State} & \textbf{Survival Prob.} & \textbf{Scaling Factor Range} \\
  \midrule
  0: Safe      & $S(t) > $ 0.55           & 1.0$\times$        \\
  1: Proactive & 0.50 $< S(t) \leq $ 0.55 & [1.1, 1.5]$\times$ \\
  2: Warning   & 0.40 $< S(t) \leq $ 0.50 & [1.3, 1.8]$\times$ \\
  3: Critical  & $S(t) \leq $ 0.40        & [1.5, 2.0]$\times$ \\
  \bottomrule
  \end{tabular}%
  }
  \vspace{-0.5em}
\end{table}

\para{KM (Kaplan-Meier) model}
\FIG~\ref{fig:spot-survival} shows the conditional survival curve for {\tt p3.2xlarge} spot instances in AWS region us-west-2a (from SkyPilot trace \cite{cantbelate.nsdi24}). 
The probability falls steeply within the first few hours, then tails off.
This time-dependent pattern motivates a survival-aware scaling policy.
\sys employs Kaplan--Meier (KM) estimator~\cite{kaplan.jasa58}, a standard nonparametric method in survival analysis, to estimate per-instance survival probability from observed uptime.
Spot preemption risk varies with uptime across instance types and regions, so no single distribution fits them all. 
The KM model instead builds the survival curve directly from each observed trace, without assuming any distribution.

For a spot instance with current uptime $u$, the KM model returns the probability that it survives an additional time window $\Delta t$:
\begin{equation}
S(\Delta t \mid u) = P(T > u + \Delta t \mid T > u),
\label{eq:km-conditional}
\end{equation}
where $T$ is the instance's total lifetime.
We write this conditional probability as $S(t)$ in the rest of the paper. 
$S(t)$ is looked up from the pre-computed curve in Figure~\ref{fig:spot-survival}.
Spot instances are then classified into survival states (\TAB~\ref{tab:survival-states}) based on $S(t)$.
Spot instances in the safe state require no intervention, while instances in `Proactive', `Warning', or `Critical' states are candidates for scale-out.
Please note that the KM model is deferred until enough preemption events are observed to fit a stable curve. 

\para{State-based scaling}
When a spot instance is classified into a non-safe survival state, the workflow running on the VM is at risk of preemption. 
\sys mitigates this risk by scaling out the workflow onto safer instances.
The remaining question is by {\em how much to scale out}.
The amount of scale-out, the {scaling factor}, is determined by the survival state: 
a higher state (closer to preemption) maps to a larger scaling factor (\TAB~\ref{tab:survival-states}). 
Each state defines a range, and within the range the factor scales based on $S(t)$ from the KM model.
Because each additional replica incurs cost, this trade-off balances preemption risk against additional spending (due to scale-out operations). 
The factor $f$ multiplies each currently running task's parallelism, and a task with minScale $m$ is expanded to $\lceil f \cdot m \rceil$ replicas, distributed across safer VMs.
For example, if a spot instance is in the `2: Warning' state ($f = 1.5$ within its [1.3, 1.8]$\times$ range), a task with $m = 4$ on the instance yields 6 replicas, which are deployed on other safer spot VMs.
If the safer-VM pool approaches capacity, \sys provisions additional VMs proactively, anticipating saturation from the current deployment plan.
Algo.~\ref{alg:scaler} details the scaling lookup and replica dispatch.

\begin{algorithm}[t]
\caption{Adaptive Scaler Decision Loop}
\label{alg:scaler}
\begin{algorithmic}[1]
\Require Spot instances $I$, uptime $u_i$ for each $i \in I$
\Procedure{AdaptiveScale}{$I$}
    \If{KM model not yet ready} \State \Return \EndIf
    \For{each spot instance $i \in I$}
        \State $S \gets \textsc{KM.lookup}(u_i)$ \Comment{Survival Prob. from KM Model}
        \State $\textit{state} \gets \textsc{classify}(S)$
        \If{$\textit{state} = \textsc{Safe}$} \State \textbf{continue} \EndIf
        \State $f \gets \textsc{lookup\_factor}(\textit{state}, S)$ \Comment{in \TAB~\ref{tab:survival-states}}
        \State $\textit{tasks} \gets \textsc{sort\_desc}(\textsc{tasks\_on}(i), \text{key} = \text{HEFT rank})$
        \For{each task $t$ in $\textit{tasks}$}
            \State $m \gets \textsc{minScale}(t)$
            \State $\textit{replicas} \gets \lceil f \cdot m \rceil$
            \State \textsc{deploy}($t$, $\textit{replicas}$, target = safer instances)
        \EndFor
        \If{safer-instance pool is saturating}
            \State \textsc{provision\_new\_vms}()
        \EndIf
    \EndFor
\EndProcedure
\end{algorithmic}
\end{algorithm}

\para{HEFT-ordered scale-out}
Scaling resources are finite. 
To maximize workflow completion under spot preemption, \sys creates a scale-out sequence for tasks on at-risk spot instances based on HEFT rank (\S\ref{subsec:dag-analysis}). 
As a result, even if scale-out is interrupted by capacity saturation, \critical tasks are more likely to have already received additional replicas. 
This HEFT-based ordering reduces the risk of cascading failures. 

\subsection{Lineage-based Retry}\label{subsec:retry}

The retry manager recovers preempted workflow tasks by resuming each from its nearest upstream checkpoint rather than restarting from the workflow's origin. 
This manager replays only the sub-DAG downstream of that recovery point, then sizes the retry batch according to how far the current success rate falls below the target. 
Algo. \ref{alg:checkpoint-retry} formalizes this decision loop.

\para{Checkpoint-based recovery}
When a task is preempted, the retry manager identifies the recovery point by traversing the DAG backward from the failed task. 
Because checkpoints are written upon task completion (in \S\ref{subsec:chkptmgr}), the recovery point is the nearest completed predecessor whose tier guarantees that a checkpoint has been persisted. 
Execution resumes from that point, replaying only the sub-DAG downstream of the recovery point rather than the full workflow. 
{Each workflow run is subject to a maximum retry limit $L_{\max}$ (Algo. \ref{alg:checkpoint-retry}, line 2), 
set between 12 and 35 by workflow size. 
A run that uses all its retries without completing is counted as a failure, which is why success rates stay below 100\%.}

\begin{table}[t]
\caption{Retry batch-sizing policy in the retry manager
  (Algo.~\ref{alg:checkpoint-retry}).
  {\normalfont \em 
  $\delta = \max\!\bigl(0,\,(\sigma_t - \sigma_c)/\sigma_t\bigr)$ is the normalized gap between the current success rate~$\sigma_c$ and the operator-specified target ~$\sigma_t$. 
  $B$~is the number of candidates retried per cycle, 
  and $\alpha$ is a multiplier applied when scale-out is inefficient. 
  {We empirically set $B_{\mathrm{high}}$, $B_{\mathrm{mid}}$, $B_{\mathrm{low}}$, and $\alpha$ to 250, 200, 150, and 1.3.}}}
\label{tab:retry-batch-policy}
\vspace{-0.5em}
\resizebox{0.9\columnwidth}{!}
{
    \begin{tabular}{@{}lll@{}}
    \toprule
    \textbf{Condition} & \textbf{Batch Size} & \textbf{Effect} \\
    \midrule
    $\delta > \delta_{\mathrm{high}}$
      & $B_{\mathrm{high}}$ & Aggressive recovery \\
    $0 < \delta \le \delta_{\mathrm{high}}$
      & $B_{\mathrm{mid}}$  & Moderate recovery \\
    $\delta = 0$
      & $B_{\mathrm{low}}$  & Maintenance only \\
    \begin{tabular}[c]{@{}l@{}}Scale-out ineffective \\ for $\ge$ 2 cycles\end{tabular}
      & $B \times \alpha$  & \begin{tabular}[c]{@{}l@{}}Shift effort from scaling \\ to retrying \end{tabular}  \\
    \bottomrule
    \end{tabular}
}
\vspace{-0.5em}
\end{table}

\begin{algorithm}[t]
\caption{Lineage-based retry}
\label{alg:checkpoint-retry}
\begin{algorithmic}[1]
\Require Incomplete workflow runs $U$, retry history, workflow success rates ($\sigma_t$, $\sigma_c$)
\Ensure Batch of runs retried from latest checkpoints

\Statex \text{// Compute success gap}
\State $\delta \gets \max\!\bigl(0,\;(\sigma_t - \sigma_c)\,/\,\sigma_t\bigr)$

\Statex \text{// Filter and rank candidates}
\State $C \gets \{u \in U : \text{incomplete}, R[u] < L_{\max}[u]\}$
\For{$u \in C$}
    \State $p(u) \gets \pi_w + \lfloor \delta \times \lambda_d \rfloor + (n_c \times w_c) - (r \times w_r)$
\EndFor
\State Sort $C$ descending by $p(u)$

\Statex \text{// Set batch size (Table~\ref{tab:retry-batch-policy})}
\If{$\delta > \delta_{\mathrm{high}}$} $B \gets B_{\mathrm{high}}$
\ElsIf{$\delta > 0$} $B \gets B_{\mathrm{mid}}$
\Else{} $B \gets B_{\mathrm{low}}$
\EndIf

\Statex \text{// Enlarge if scale-out is ineffective}
\If{scale-out has not improved $\sigma_c$ for $\ge 2$ cycles}
    \State $B \gets \lfloor B \times \alpha \rfloor$
\EndIf

\Statex \text{// Retry from latest checkpoint}
\For{top $B$ runs from sorted $C$}
    \State Resume from latest checkpoint
    \State Increment retry count
\EndFor

\end{algorithmic}
\end{algorithm}

\para{Retry batch sizing}
The checkpoint-based recovery (described above) determines how to resume a single failed run. 
In practice, preemptions in a shared spot pool can leave multiple workflow runs incomplete simultaneously.
However, immediately retrying all incomplete runs is undesirable, as they will compete for the same limited spot capacity.
Therefore, the retry manager selects a subset of candidates to retry each cycle. 
From all incomplete runs, only those below their maximum retry limit are eligible. 
Each eligible workflow run~$u$ receives a priority score:
\begin{equation}
\label{eq:priority-score}
p(u) = \pi_w + \lfloor \delta \times \lambda_d \rfloor
     + (n_c \times w_c) - (r \times w_r),
\end{equation}
where $\pi_w$ is a per-workflow base priority,
$\lambda_d$ is an urgency scaling constant,
$n_c$ is the number of tasks already completed in $u$,
$w_c$ is the per-task completion weight,
$r$ is the number of prior retries for $u$,
$w_r$ is the per-retry penalty weight,
and $\delta$ is the normalized success-rate gap defined in \TAB\ref{tab:retry-batch-policy}.

The score increases with the number of completed tasks and decreases with the number of prior retries, so that near-complete runs are retried first.
{Tier enters this ordering only indirectly, since resuming from a higher-tier checkpoint preserves more completed}
{tasks. Because each retry queue holds runs of one workflow, normalizing $n_c$ by workflow size would not change the retry order.}
The operator specifies a target success rate~$\sigma_t$ for the workflow, and the retry manager uses the gap~$\delta$ between~$\sigma_t$ and the current rate~$\sigma_c$ to control the batch size (\TAB\ref{tab:retry-batch-policy}). 
Therefore, a large gap triggers aggressive recovery, while a small or zero gap reduces the batch to conserve spot capacity
(Algo.~\ref{alg:checkpoint-retry}, lines~6--8). 
When the recent scale-out operations of the adaptive scaler (\S\ref{subsec:scaler}) fail to improve the success rate for two or more consecutive cycles, the retry manager enlarges the batch further, shifting recovery effort from scaling to retrying (Algo.~\ref{alg:checkpoint-retry}, lines~9--10).

\section{Implementation}\label{sec:Checkscale Implementation}

\subsection{\sys Implementation}

\sys is implemented in approximately 17.5K lines of Python on a Kubernetes cluster using Fission as the serverless execution platform. 
Fission was selected over other open-source FaaS frameworks for its native workflow support and ease of deployment. 
All runtime state (\eg, checkpoint records, retry history, and scaling decisions) is persisted in MySQL. 
{Checkpoint state is written to the same store, which runs on a non-spot VM so that a preemption cannot erase it. 
Writes are transactional, so a failure during a checkpoint write leaves the previous checkpoint intact and recovery falls back to it.}
The offline analysis outputs, $\kappa(v)$ and $\ell(v)$, are distributed to online controllers via per-task Kubernetes {\tt ConfigMaps} populated before workflow deployment.

{\em Task ranker} runs as a standalone process at deployment time. 
This component parses the workflow DAG definition ({\tt tasks.json}), computes $\kappa(v)$ for each task, and writes the result to a local file used by the checkpoint planner.

{\em Checkpoint planner} runs after the task ranker completes. 
This planner reads $\kappa(v)$ and the DAG definition, assigns $\ell(v)$ for each task, and populates per-task Kubernetes {\tt ConfigMaps} via the cluster API. 
At runtime, the checkpointing mechanism embedded in each task reads $\ell(v)$ from its attached {\tt ConfigMap} and writes task output state to MySQL at the configured interval.

{\em Adaptive scaler} runs as an independent subprocess on the cluster control plane. 
At startup, this component generates a KM survival curve from spot preemption traces and loads it into memory. 
At each polling interval, it retrieves active pod uptimes, evaluates survival probability against the curve, and issues scale-out operations prioritized by $\kappa(v)$. 
Scaling decisions are recorded in MySQL for coordination with the retry manager.

{\em Retry manager} runs as an independent subprocess alongside the scaler. 
This manager polls MySQL at regular intervals to detect failed tasks and dispatches recovery requests to the Fission gateway. 
Prior retry history is read at startup to calibrate batch sizing, and retry decisions are persisted to the same store.

The \sys implementation is available at
{\urlstyle{tt}\href{https://github.com/bandhanp1804/faas-spot}%
{\nolinkurl{https://github.com/bandhanp1804/faas-spot}}}.

\subsection{Trace-driven Spot Emulator}\label{subsec:spot-emul}

Evaluating \sys on live cloud spot instances is impractical for two reasons. 
Spot market conditions shift continuously, so back-to-back runs of different systems face different interruption sequences, making a fair paired comparison impossible. 
Moreover, sustained evaluation across multiple workflows and methods at scale is financially expensive and time-consuming.

Therefore, we developed a trace-driven spot emulator to fairly evaluate the performance of \sys and competing approaches. The emulator is designed around three goals. 
\emph{1) Trace fidelity}: we used spot preemption events that are derived from real production traces. 
The emulator replays empirical lifetime distributions collected from spot markets \cite{lee.iiswc22, cantbelate.nsdi24}. 
\emph{2) Fair evaluation}: all methods under comparison receive identical spot preemption events within each trial, so reported differences reflect controller quality rather than differences in the underlying spot conditions. 
\emph{3) Controller-agnostic design}: we developed the emulator to be decoupled from \sys and expose an evaluation API, which allows any spot-aware system to plug in as an evaluation target.

\para{Spot traces}
The emulator uses two public trace sources. 
Spot instance lifetimes are derived from the SkyPilot availability traces~\cite{cantbelate.nsdi24}, which record uptime across multiple spot VMs over a two-week period with different regions, operating systems, and instance types. 
The emulator fits an empirical lifetime distribution from these traces.
\FIG\ref{fig:spot-survival} (in \S\ref{sec:system_components}) shows the resulting survival curve for {\tt p3.2xlarge} VMs in AWS us-west-2a region. 
Spot pricing is obtained from SpotLake~\cite{lee.iiswc22}, which archives historical spot prices across regions and instance types. 
The emulator applies these prices to compute per-VM execution cost over each VM's observed lifetime.

\para{Spot emulator architecture}
The emulator consists of four components.
The \emph{distribution extractor} derives an empirical lifetime
distribution from the raw traces, filtered by region, operating
system, and instance type, as lifetime characteristics vary
significantly across these dimensions~\cite{spotcheck.eurosys15, cloudindex.socc18, cantbelate.nsdi24}.
The \emph{lifecycle manager} emulates the lifecycle of each spot VM. 
When a new VM is provisioned, it samples a lifetime between $p_{10}$ and $p_{90}$ of the derived distribution, excluding extreme outliers at both tails. 
The manager runs at 60$\times$ accelerated time, enabling large-scale workflow evaluation without the time and cost overhead of live spot experiments. 
It tracks each VM's uptime on this accelerated clock and triggers a preemption event when the uptime reaches the assigned lifetime.
The \emph{cost calculator} computes the execution cost for each VM using the historical spot prices obtained from SpotLake. 
When a VM is preempted or completes its workload, the calculator records the accrued cost over the VM's observed lifetime.
The \emph{preemption dispatcher} provides the interface between the emulator and the execution plane. When a VM's uptime reaches its assigned lifetime, the dispatcher deletes the affected instance, terminates all co-located tasks, and notifies the cost calculator to record the accrued cost.

\para{Execution model}
The emulator runs on a physical server cluster. Each physical machine hosts a VM with dedicated cores and memory, and Kubernetes runs across these VMs with each pod representing a spot VM. The lifecycle manager assigns each pod a lifetime sampled from the derived distribution and preempts it when its uptime reaches that lifetime. Workflow tasks run as containers inside these pods via Fission, so workflows execute on real infrastructure under emulated preemption. A preemption terminates all containers in the affected pod, matching instance-level preemption at real providers. A fixed seed assigns pod lifetimes in each trial, so all methods receive identical interruption sequences within a paired trial.

\section{Evaluation}\label{sec:eval-v1}

We evaluate \sys on the trace-driven spot emulator (\S\ref{subsec:spot-emul}) with \wfcnt structurally diverse DAG workflows and four baselines. 
The key results are:
\begin{itemize}[leftmargin=1em]
  \item \sys achieves 89.03\% average success rate at \$0.018 cost per success, outperforming all four baselines on both metrics simultaneously (\S\ref{subsec:eval-succ-cost}).
  \item \sys's mean latency of 126s is 2.3$\times$ lower than the nearest competitors, with the smallest cross-workflow variance at every percentile (\S\ref{subsec:eval-latency}).
  \item \sys writes 10.6 checkpoints per successful run, while baselines write 1.17$\times$ to 3.8$\times$ more, because it places checkpoints selectively at cascade-prone positions (\S\ref{subsec:eval-checkpoint}).
  \item Ablation confirms that all three fault-tolerance mechanisms are essential, as removing any one increases cost per success by 1.8$\times$ to 4.6$\times$, and removing lineage-based retry reduces the success rate to 50\% (\S\ref{subsec:eval-ablation}).
\end{itemize}

\subsection{Evaluation Setup}

\begin{figure}[t]
  \centering
  \includegraphics[width=1\linewidth]{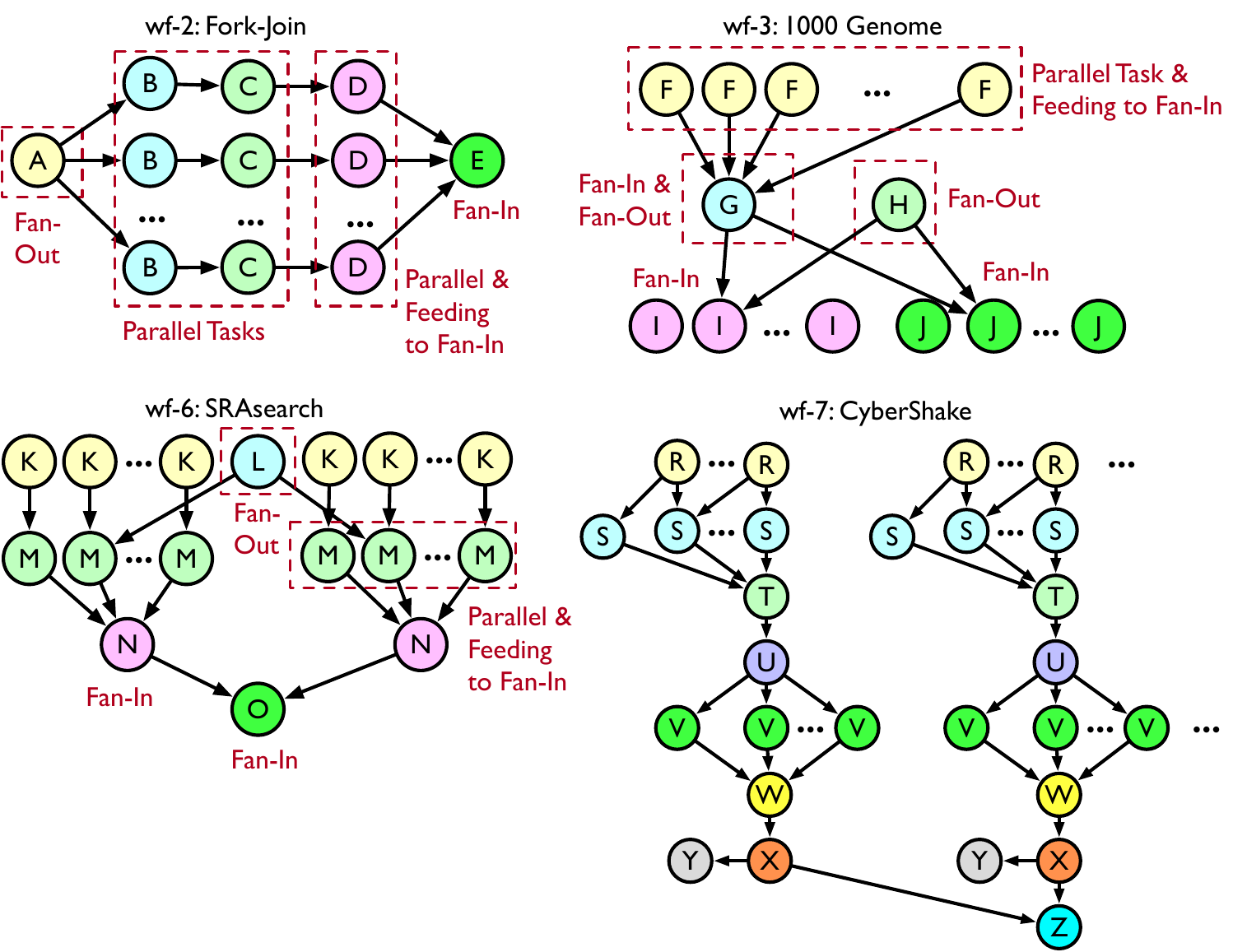}
  \vspace{-0.5em}
  \caption{DAG structures of four additional workflows used in evaluation.
  {\normalfont \em `wf-2': a fork-join DAG, `wf-3': 1000Genome \cite{mashup.ppopp22, 1000genomes.nature}, `wf-6': SRAsearch \cite{mashup.ppopp22, gridgreen.socc25}, and `wf-7': CyberShake workflow \cite{workflows.works08}.}}
  \label{fig:wf-suite}
  \vspace{-1em}
\end{figure}

\para{Serverless workflow suite}
We use eight DAG workflows drawn from four application domains. 
Three workflows with high cascade potential (wf-4, wf-5, wf-8) were introduced in \S\ref{sec:motivation} (\FIG\ref{fig:three-wfs}). 
\FIG\ref{fig:wf-suite} shows four additional workflows: wf-2, a fork-join DAG with a single fan-out/fan-in pair; 
wf-3 (1000Genome \cite{mashup.ppopp22, 1000genomes.nature}), a pipeline with large fan-out/fan-in stages; 
wf-6 (SRAsearch \cite{mashup.ppopp22, gridgreen.socc25}), a two-branch DAG with parallel tasks per branch; 
and wf-7 (CyberShake \cite{workflows.works08}), a wide DAG with repeated fan-out/fan-in stages. 
The remaining workflow, wf-1, is a compact DAG with a single fan-out and no synchronization barrier.

\para{Baselines}
\sys is evaluated against four baselines. 
Snape \cite{snape.asplos23} dynamically adjusts the ratio of spot to on-demand instances using eviction prediction and constrained reinforcement learning. 
Hourglass \cite{hourglass.eurosys19} manages temporal slack as a risk budget for time-constrained graph-processing jobs, falling back to on-demand instances when slack is exhausted. 
MScheduler \cite{mscheduler.cloudcom23} recovers spot-preempted HPC jobs through application-native restart and cost-based instance selection across regions. 
Burst-HADS \cite{teylo.tcc23} schedules bag-of-tasks workloads across spot and burstable instances with deadline constraints. None of the four baselines incorporates DAG structure into checkpoint placement or retry ordering.

\para{Experiment configuration and metrics}
Each workflow is executed 500 times per method per trial, and each experiment is repeated over 5 trials. 
\sys and four baselines receive identical spot preemption sequences within each trial to ensure that performance differences reflect controller behavior rather than differences in interruption conditions. 
All methods share the same DAG definitions, resource pool, and spot prices.
The experiments use three spot instance types ({\tt m3.2xlarge}, {\tt p3.2xlarge}, {\tt t3.2xlarge}) across two AWS regions (us-west-2a, us-west-2b).

We report three metrics.
Success rate is the fraction of submitted runs that complete. 
Cost per success is the total cost divided by the number of completed runs. 
{Note that the reported cost covers compute only. 
The shared store runs as a single VM serving every method, so its cost does not scale with checkpoint frequency and will not change the relative comparison.}
Average latency is the mean end-to-end wall-clock time over completed runs, including queueing, retries, and checkpoint replay. 
{Serverless function startup overhead is also included in this latency metric, as each task runs as a real pod on Kubernetes.}
Latency values are used for relative comparison only, as absolute values reflect the emulated timeline.

\begin{figure*}[t]
  \centering
  \subfloat[Workflow success rate]{%
    \includegraphics[width=0.45\linewidth]{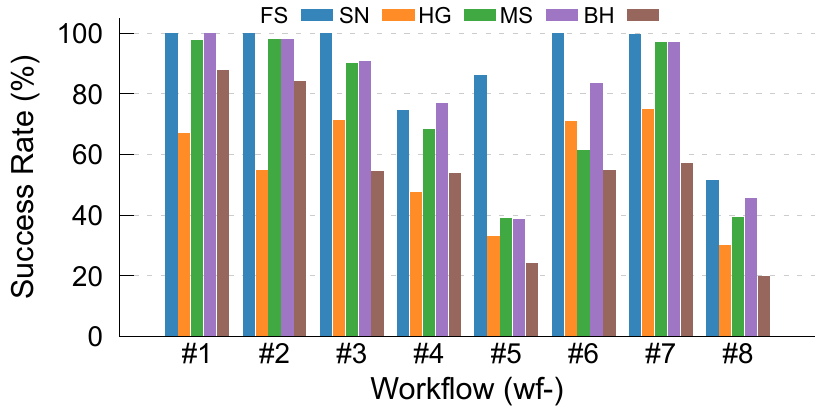}%
    \label{fig:success-rate}}%
  \hfill
  \subfloat[Cost per successful run]{%
    \includegraphics[width=0.45\linewidth]{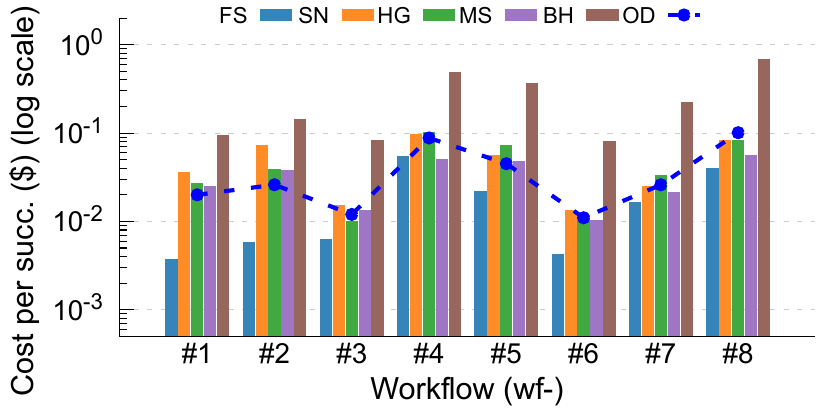}%
    \label{fig:cost-overview}}%
  \vspace{-0.5em}
  \caption{Success rate and cost per success for each workflow and baseline.
  {\normalfont \em
  {(a) Success rate across eight workflows: \sys achieves the highest success rate on seven of \wfcnt workflows, with the largest margin on wf-5 (+43 pp over the nearest baseline). 
  On-demand completes every run.}
  {(b) Cost per successful run (log scale, lower is better): 
    \sys achieves the lowest cost per success on seven of \wfcnt workflows. 
    The dashed line marks on-demand cost per successful run (\FIG \ref{fig:wf-cost}).
    `FS': \sys, `SN': Snape, `HG': Hourglass, `MS': MScheduler, and `BH': Burst-HADS.}}}
  \label{fig:eval-success-cost}
  \vspace{-0.5em}
\end{figure*}

\subsection{Success Rate and Cost Efficiency}\label{subsec:eval-succ-cost}
We report the workflow success rate and cost per successful run together, since high completion is meaningful only if it does not require excessive VM execution cost. 
\FIG\ref{fig:eval-success-cost} shows both metrics for \sys and all four baselines across the \wfcnt workflows: \FIG~\ref{fig:success-rate} for success rate and \ref{fig:cost-overview} for cost per successful run.

\sys outperforms all four baselines on both average success rate and cost per success. 
It achieves 89.03\% average success rate at only \$0.018 cost per success. 
The closest competitor, MScheduler, reaches 81.78\% at \$0.04, which is 7.25 pp lower in success rate and more than 2$\times$ higher in cost per completed run. 
Hourglass reaches 76.22\% at \$0.047, Snape 58.56\% at \$0.05, 
and Burst-HADS 53.43\% at \$0.259 (\TAB\ref{tab:aggregate-results}).

\begin{figure}[t]
  \centering
  \includegraphics[width=0.9\linewidth]{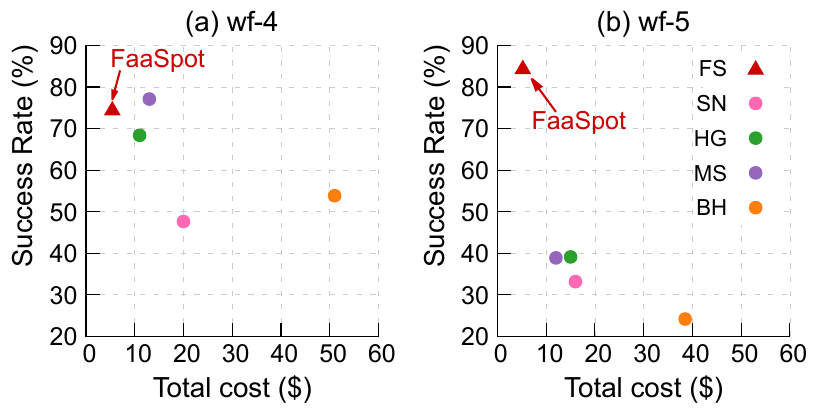}
  \vspace{-0.5em}
  \caption{Total cost vs. success rate for wf-4 and wf-5. 
  {\normalfont \em 
  The top-left region indicates higher success at lower cost. 
  On wf-4, \sys (red triangle) has the lowest cost while MScheduler reaches slightly higher success at over twice the cost. 
  On wf-5, \sys leads on both axes with all four baselines clustering at higher cost and lower success.}}
  \label{fig:pareto}
  \vspace{-0.5em}
\end{figure}

The differences in success rate and cost vary with workflow DAG structure. 
We observe that, on wf-4, MScheduler surpasses \sys by 2.5 pp on success (77.1\% vs. 74.6\%), but at\ \$0.067 per success vs. \sys's \$0.055 (22\%$\downarrow$). 
On wf-5, \sys reaches 86.2\% while all four baselines fall below 41\%.
\FIG\ref{fig:pareto} plots total cost against success rate for these two high-cascade workflows. 
We omit the remaining workflows for brevity, but they show a similar pattern.

\begin{table}[t]
  \centering
  \small
  \caption{Average success rate and cost per success across the \wfcnt test workflows. 
  {\normalfont \em Best result per column in bold.}}
  \label{tab:aggregate-results}
  \vspace{-0.5em}
  \begin{tabular}{lcc}
    \toprule
    \textbf{Method} & \textbf{Success Rate (\%)} & \textbf{Cost/Succ. Run (\$)} \\
    \midrule
    {\bf \sys}          & \textbf{89.03\,$\pm$\,5.87} & \textbf{0.018\,$\pm$\,0.007} \\
    Snape       & 58.56\,$\pm$\,6.17  & 0.050\,$\pm$\,0.011 \\
    Hourglass   & 76.22\,$\pm$\,8.39  & 0.047\,$\pm$\,0.011 \\
    MScheduler  & 81.78\,$\pm$\,7.95  & 0.040\,$\pm$\,0.007 \\
    Burst-HADS  & 53.43\,$\pm$\,7.74  & 0.259\,$\pm$\,0.071 \\
    \bottomrule
  \end{tabular}
  \vspace{-0.5em}
\end{table}

The performance gap is primarily due to the four baselines not considering the DAG structure when making fault-tolerance decisions.
They treat a preemption on a sequential path and one at a cascade-prone position identically. 
In contrast, \sys's offline tier assignment (\S\ref{subsec:dag-analysis}) distinguishes these cases before the first task executes, placing checkpoints only at cascade-prone positions. 
When a preemption occurs at a synchronization barrier, recovery focuses on the sub-DAG downstream of the nearest checkpoint rather than requiring a full-workflow restart.

\subsection{End-to-end Latency}\label{subsec:eval-latency}

We next report end-to-end latency, and \TAB\ref{tab:tail-latency} shows the mean, p50, p95, and p99 latency for each method. 
\sys achieves the lowest latency across all four percentiles. 
Its mean of 126s is 2.3$\times$ lower than the nearest competitors, which are Snape (297s) and Burst-HADS (290s). 
The latency difference is especially large at the tail.
At p95, \sys achieves 245s, compared to 704s for Snape, 635s for MScheduler, and 1310s for Burst-HADS. 
At p99, the gap increases further, with Hourglass reaching 2199s and Burst-HADS 1897s against \sys's 346s.

\begin{table}[htbp]
  \centering
  \small
  \caption{End-to-end latency across the eight test workflows (sec). 
  {\normalfont \em Each column reports the workflow-averaged percentile over completed runs, with $\pm$ indicating cross-workflow variation. 
  }}
  \label{tab:tail-latency}
  \vspace{-0.5em}
  \begin{tabular}{lcccc}
    \toprule
    \textbf{Method} & \textbf{Mean} & \textbf{p50} & \textbf{p95} & \textbf{p99} \\
    \midrule
    {\bf \sys}          & {\bf 126\,$\pm$\,47} & \textbf{110\,$\pm$\,42}  & \textbf{245\,$\pm$\,85}   & \textbf{346\,$\pm$\,108}  \\
    Snape         & 297\,$\pm$\,108 & 264\,$\pm$\,114          & 704\,$\pm$\,223           & 799\,$\pm$\,237           \\
    Hourglass     & 506\,$\pm$\,139 & 443\,$\pm$\,143          & 1138\,$\pm$\,339          & 2199\,$\pm$\,219          \\
    MScheduler    & 334\,$\pm$\,128 & 326\,$\pm$\,130          & 635\,$\pm$\,237           & 680\,$\pm$\,250           \\
    Burst-HADS    & 290\,$\pm$\,66 & 115\,$\pm$\,64           & 1310\,$\pm$\,158          & 1897\,$\pm$\,209          \\
    \bottomrule
  \end{tabular}
  \vspace{-0.5em}
\end{table}

Beyond absolute values, \sys also shows the smallest variance across workflows. Its $\pm$ values in \TAB\ref{tab:tail-latency} are consistently the lowest (\eg, $\pm$42 at p50, compared to $\pm$114 for Snape and $\pm$143 for Hourglass), indicating that \sys delivers predictable latency.
The baselines show high variance because their recovery cost depends on where in the DAG the preemption happens to land. 
In contrast, \sys's lineage-based retry (\S\ref{subsec:retry}) reduces this variance by always resuming from the nearest upstream checkpoint rather than restarting from the workflow origin.

\subsection{Checkpoint Efficiency}\label{subsec:eval-checkpoint}
\begin{figure}[t]
  \centering
  \includegraphics[width=0.9\linewidth]{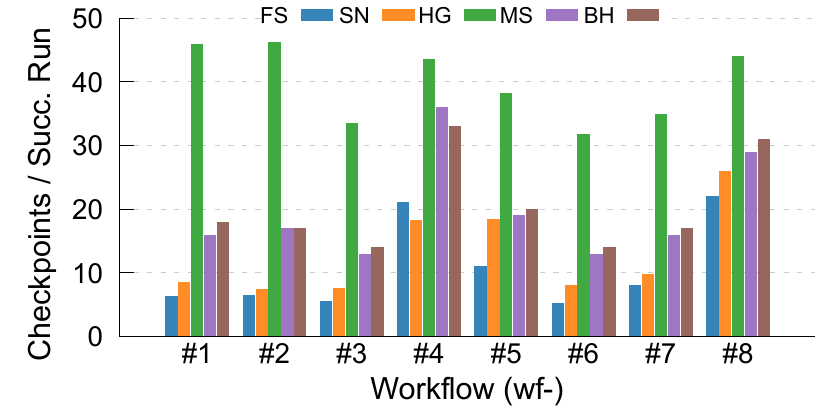}
  \vspace{-0.5em}
  \caption{Average number of checkpoints per successful run. 
  {\normalfont \em Lower values indicate more selective checkpoint placement. 
  \sys achieves the fewest checkpoints per success on every workflow.}}
  \label{fig:ckpt-diagnostics}
  \vspace{-0.5em}
\end{figure}

We next report checkpoint efficiency by measuring the number of checkpoints each method requires for successful completion. 
A lower count indicates that the method places checkpoints more selectively, reducing I/O overhead without sacrificing reliability. 
\FIG\ref{fig:ckpt-diagnostics} reports the average number of checkpoints per successful run for each method across the eight workflows.

\sys completes workflow executions with an average of 10.6 checkpoints per successful execution, the fewest among all evaluated methods. 
Snape requires 12.4 (1.17$\times$ $\uparrow$), MScheduler 19.6 (1.8$\times$ $\uparrow$), and Hourglass 39.9 (3.8$\times$ $\uparrow$) checkpoints per successful execution.
Among the baselines, Snape requires the fewest checkpoints per success (12.4 vs. \sys's 10.6), but this is because Snape partially uses on-demand instances, so fewer tasks are exposed to preemption in the first place. 
None of the baselines selects checkpoint positions based on the DAG structure. Hourglass checkpoints every task uniformly, which is the main reason for its highest checkpoint count (39.9) among all methods.
\sys's tier assignment $\ell(v)$ concentrates checkpoints at fan-out nodes and tasks feeding fan-in nodes, the positions where a missed checkpoint forces the most downstream re-execution.

\subsection{Ablation Study}\label{subsec:eval-ablation}

\begin{table}[t]
  \centering
  \small
  \caption{Contribution of each mechanism. 
  {\normalfont \em Average success rate and cost per successful run for \sys and three ablated variants, each removing one mechanism. Best value per column in bold.}}
  \label{tab:abl-results}
  \vspace{-0.5em}
  \resizebox{0.9\columnwidth}{!}
  {
    \begin{tabular}{lcc}
      \toprule
      \textbf{Variant} & \textbf{Avg. Succ. (\%)} & \textbf{Avg. Cost/Succ. (\$)} \\
      \midrule
      {\bf \sys}                    & \textbf{89.03}  & \textbf{0.018} \\
      A (w/o Selective Ckpt.)    & 86.44           & 0.033 \\
      B (w/o Adapt. Scaling)        & 86.74           & 0.041 \\
      C (w/o Lineage Retry)        & 50.05           & 0.082 \\
      \bottomrule
    \end{tabular}
  }
  \vspace{-0.5em}
\end{table}

To isolate the contribution of each mechanism, we evaluate three ablated variants of \sys. 
Each variant removes one mechanism while keeping the others. 
`Variant-A' removes selective checkpointing and instead checkpoints every task uniformly, as in systems that do not consider DAG structure. 
`Variant-B' removes survival-aware scaling, so tasks are recovered through checkpointing and retry alone, without proactive scale-out onto safer instances. 
`Variant-C' removes lineage-based retry. 
As checkpoints are used exclusively for lineage-based recovery, checkpointing is also disabled. 
Consequently, a preempted workflow is restarted from its origin. 
We compare all three against full \sys on the \wfcnt-workflow suite, reporting success rate and cost per successful run.

\TAB\ref{tab:abl-results} reports the average success rate and cost per successful run for each variant across the eight workflows. The three variants reveal a consistent pattern. 
Removing selective checkpointing (Variant-A) or survival-aware scaling (Variant-B) causes only a modest reduction in success rate (86.44\% and 86.74\%, respectively, vs. 89.03\% for \sys), but substantially increases the cost of achieving a successful run. 
In contrast, removing lineage-based retry (Variant-C) reduces the success rate to 50.05\% and increases the cost per successful run to \$0.082.

The three variants isolate where each mechanism's benefit comes from. 
Variant-A checkpoints every task and achieves a success rate similar to that of \sys. 
However, many of these checkpoints occur at sequential positions where re-execution is inexpensive, increasing the cost to \$0.033. 
This result shows that selective checkpointing improves efficiency not by increasing the success rate, but by avoiding checkpoints where the benefits of recovery are small.
Variant-B recovers without survival-aware scaling, so recovery relies entirely on checkpointing and retry. 
As a result, more tasks must be re-executed after spot preemptions, which naturally increases the cost to \$0.041, despite a success rate close to \sys's.
Variant-C shows the largest degradation.
Without checkpoint-based recovery, every preemption forces the workflow to restart from its origin, reducing the success rate to 50.05\% and increasing the cost to \$0.082. 
These results show that lineage-based retry is the primary contributor to recovery effectiveness, and both selective checkpointing and survival-aware scaling improve the efficiency of that recovery. Together, the three mechanisms enable \sys to achieve both high completion and low cost on spot instances.

\begin{figure}[t]
  \centering
  \subfloat[Workflow success rate]{%
    \includegraphics[width=0.49\linewidth]{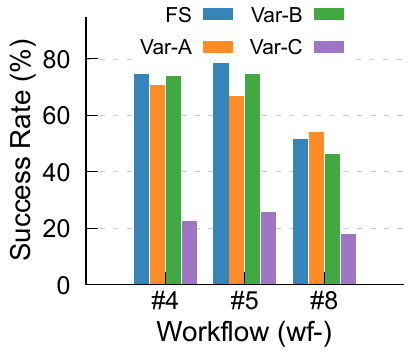}%
    \label{fig:abl-succ}}%
  \hfill
  \subfloat[Cost per successful run]{%
    \includegraphics[width=0.49\linewidth]{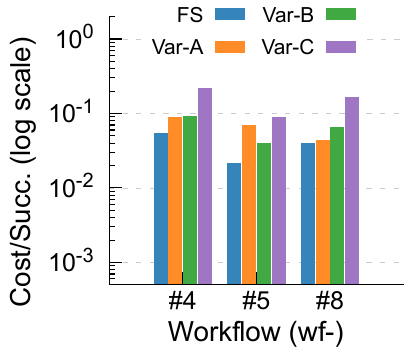}%
    \label{fig:abl-cost}}%
  \vspace{-0.5em}
  \caption{Ablation results on the three high-cascade workflows (wf-4, wf-5, wf-8).
  {\normalfont \em 
  (a) Success rate and 
  (b) cost per successful run (log scale) for \sys (FS) and three variants: Var-A (w/o selective checkpointing), Var-B (w/o survival-aware scaling), and Var-C (w/o lineage-based retry).}}
  \label{fig:abl-wf-458}
  \vspace{-0.5em}
\end{figure}

We now take a closer look at the three high-cascade workflows: wf-4, wf-5, and wf-8. 
The results are shown in \FIG\ref{fig:abl-wf-458}.
Removing checkpoint-based recovery (Variant-C) shows the largest degradation here, reducing success rates to 22.7\%, 25.9\%, and 17.9\%, respectively.
This degraded performance is mainly due to each preemption discarding all prior progress, forcing these long, highly parallel workflows to restart from their origins. 
On wf-8, where every variant performs poorly,
Variant-A achieves a slightly higher success rate than \sys (54.2\% vs. 51.7\%) because checkpointing every task recovers a few additional runs. 
However, this improvement comes at a higher cost per successful run (\$0.044 vs. \$0.040). 
This is the trade-off selective checkpointing is designed to make. 
\sys skips checkpoints where the benefit is small, accepting a slightly lower completion rate in return.

\begin{table}[b]
\vspace{-1.0em}
\caption{
{{Sensitivity to the survival state boundaries (Table~\ref{tab:survival-states}) and the retry batch size (Table~\ref{tab:retry-batch-policy}).} 
{\normalfont \em  For the boundaries, all three shift together, and the safe boundary is shown. Each cell reports the success rate (\%) and cost per successful run (\$) for wf-4/5/8 in \FIG\ref{fig:three-wfs} (in \S\ref{sec:motivation}). We used the default settings (in bold) in our evaluation.}}}
\label{tab:param-sensitivity}
\vspace{-0.5em}
\centering
\small
\resizebox{0.95\columnwidth}{!}
{
  \begin{tabular}{lccc}
    \toprule
    \textbf{Configuration} & \textbf{wf-4} & \textbf{wf-5} & \textbf{wf-8} \\
    \midrule
    \multicolumn{4}{l}{{\em Survival state boundaries}} \\
    \midrule
    Safe at {\em S(t)} = 0.45 & 50.2\% / \$0.12 & 56.7\% / \$0.05 & 27.0\% / \$0.09 \\
    Safe at {\em S(t)} = 0.50 & 62.8\% / \$0.08 & 69.2\% / \$0.04 & 36.0\% / \$0.06 \\
    {\bf Safe at {\em S(t)} = 0.55} & \textbf{74.6\% / \$0.06} & \textbf{86.2\% / \$0.03} & \textbf{51.7\% / \$0.04} \\
    \midrule
    \multicolumn{4}{l}{{\em Retry batch size:} ($B_{\mathrm{high}}$, $B_{\mathrm{mid}}$, $B_{\mathrm{low}}$)} \\
    \midrule 
    (150, 100, 50) & 68.6\% / \$0.07 & 77.5\% / \$0.05 & 49.0\% / \$0.04 \\
    {\bf (250, 200, 150)} & \textbf{74.6\% / \$0.06} & \textbf{86.2\% / \$0.03} & \textbf{51.7\% / \$0.04} \\
    (350, 300, 250) & 72.2\% / \$0.08 & 89.2\% / \$0.06 & 56.6\% / \$0.08 \\
    \bottomrule
  \end{tabular}
}
\end{table}

\subsection{{Parameter Sensitivity}}\label{subsec:eval-sensitivity}

{In our evaluation, we used the `safe boundary' and `batch sizes' specified in Tables \ref{tab:survival-states} and \ref{tab:retry-batch-policy}. 
We test how sensitive the results are to both parameters, using wf-4, 5, and 8. 
Table \ref{tab:param-sensitivity} reports the results.}

{The results confirm that the survival-state boundaries significantly affect both metrics. 
We shifted all three boundaries down together so that the scaler intervenes at a lower survival probability. 
Moving this safe boundary from 0.55 to 0.45 degrades the success rate on wf-8 from 51.7\% to 27.0\% and doubles the execution cost.}

{A lower boundary means the scaler acts later, so a preemption is more likely to have already occurred. 
Increasing the survival boundaries raises the success rate. 
However, this change naturally increases cost per successful run, 
since replicas are added to instances that were less likely to be preempted.
The default setting therefore balances these two cases, aiming for the scaler to act early enough to prevent cascades without wasting capacity.}

{We also observe that the retry batch size has a smaller effect. 
The smaller (\eg, $B_{\mathrm{high}}$=150) batch is worse on both metrics for every workflow. 
The larger batch (\eg, $B_{\mathrm{high}}$=350) raises the success rate on wf-5 and wf-8, but increases the cost per successful run by 2$\times$ on both. 
The default setting ($B_{\mathrm{high}}$=250, $B_{\mathrm{mid}}$=200, $B_{\mathrm{low}}$=150) balances the two, reaching a comparable success rate at half the cost.}

\section{Limitations}\label{sec:limit}

We discuss some limitations of the current evaluation and \sys design.
First, we conduct all experiments on a trace-driven spot emulator, not on live cloud infrastructure. 
While the emulator ensures fair comparison by exposing all methods to identical preemption events, the results reflect relative performance differences between methods rather than absolute production-level measurements.

Second, checkpoint I/O latency is not modeled in the emulator. 
Since \sys writes fewer checkpoints than all baselines (\S\ref{subsec:eval-checkpoint}), including this overhead would likely penalize methods with higher checkpoint frequency, further strengthening \sys's advantage.

Third, the offline tier assignment assumes that the workflow DAG structure is fully known at submission time. Some scientific workflows have data-dependent branching where execution paths vary across runs (\eg, DayDream \cite{daydream.sc22}). 
Such workflows fall outside our current scope. 
Extending the offline analysis with online DAG refinement is a natural direction for future work.

{Fourth, the task ranker requires a per-task execution time estimate. 
Inaccurate estimates can shift the criticality score. 
As the scaler uses the score only to decide which tasks to scale out first, an error matters only if it changes that order. 
The tier assignment, however, uses structure alone, so a long-running task on a sequential path receives no checkpoint even when its re-execution cost is high. 
Extending it with this estimate is our future work.}

Finally, the evaluation uses three instance types ({\tt m3.2xlarge}, {\tt p3.2xlarge}, {\tt t3.2xlarge}) across two AWS regions (us-west-2a/2b). 
Spot preemption-rates differ across regions and instance families.
Their distribution shape also depends on the underlying availability mechanism.
Therefore, the KM survival estimation model and scaling thresholds (\TAB\ref{tab:survival-states}, \S\ref{subsec:scaler}) may require recalibration for deployment targets.
{This recalibration affects only the adaptive scaler, as checkpointing and retry depend on DAG structure.}

\section{Related Work}\label{sec:related_work}

\para{Fault-tolerant execution on spot instances}
{Checkpoint-based fault tolerance has been studied for decades, with early work establishing the optimal checkpoint interval \cite{young.cacm74}. 
It has been revisited extensively for spot and transient cloud resources, where instances can be revoked with short notice. 
Prior work spans task replication  \cite{poola.taas16}, spot market characterization \cite{benyehuda.teac13, spotcheck.eurosys15, cantbelate.nsdi24}, and trace-driven availability modeling \cite{lee.iiswc22, cloudindex.socc18, spoteventvisible.www24}.}

Building on this foundation, several systems address spot preemption at the resource management level.
{Hourglass \cite{hourglass.eurosys19} manages temporal slack as a risk budget for graph processing jobs,
Snape \cite{snape.asplos23} adjusts the spot-to-on-demand ratio using eviction prediction, 
Burst-HADS \cite{teylo.tcc23} schedules bag-of-tasks workloads across spot and burstable instances, and
MScheduler \cite{mscheduler.cloudcom23} leverages application-native restart with cost-based instance selection.}
All four fall back to on-demand capacity when spot becomes unreliable.
SpotVerse \cite{spotverse.middleware24} reduces cost by selecting cloud regions based on interruption frequency and placement score, but operates at the granularity of whole workflows rather than the internal DAG structure.
{TR-Spark \cite{trspark.socc16} applies lineage-aware checkpointing to Spark's stage-based DAG model, showing that structure-aware placement outperforms uniform policies on transient resources. 
Pado \cite{pado.eurosys17} likewise analyzes the job DAG, placing operators whose eviction would trigger the most recomputation on reserved nodes.}
More broadly, Flint \cite{flint.eurosys16} and portfolio-driven management \cite{sharma.sigmetrics17} adapt batch analytics to transient server pools, and Tributary \cite{tributary.atc18} provides SLO-constrained elastic control under spot preemption.

{Most prior systems place checkpoints per task without reference to downstream dependencies, and do not distinguish a preemption on a sequential path from one at a fan-out node feeding a synchronization barrier.}
TR-Spark and Pado are the closest to \sys in exploiting DAG structure to protect the positions whose loss is most costly. 
However, TR-Spark operates within Spark's synchronous stage model, and Pado depends on reserved nodes that are always available. 
Neither targets serverless workflows running on spot instances without an on-demand fallback, where tasks must be protected through checkpointing, scaling, and retry rather than placement on safe nodes.

\para{Serverless workflow execution}
{Serverless computing is widely used to run scientific and data-intensive DAG workflows. 
Wukong \cite{wukong.socc20} and Speedo \cite{speedo.socc21} provide scalable dispatch frameworks for parallel workflows. 
Mashup \cite{mashup.ppopp22} executes HPC DAG workflows through hybrid VM-serverless deployment, and 
DayDream \cite{daydream.sc22} extends this direction to dynamic scientific workflows whose execution paths vary across runs. 
FaaS-based execution of HPC workflows is further studied in HyperFlow \cite{hyperflow.fgcs20}.  
ORION \cite{orion.osdi22} and WiseFuse \cite{wisefuse.sigmetrics22} optimize serverless DAG execution via DAG-aware resource sizing and function bundling. 
Jolteon \cite{jolton.nsdi24} configures resources under a latency or cost constraint, 
Dexter \cite{dexter.middleware24} tunes per-stage parallelism,  
MinFlow \cite{minflow.fast24} reduces the data-passing overhead of DAG shuffles, and
Netherite \cite{netherite.vldb22} shows that state persistence and checkpointing are central to efficient serverless workflow execution. 
On the platform side, prior work has been developed to optimize function-level resource allocation, cold start, latency-sensitive serving, and provider-side management \cite{bilal.eurosys23, pronghorn.eurosys24, nightcore.asplos21, trident.tsc25}.}

However, none of these studies combines serverless DAG execution with transient spot instance pools, leaving the fault-tolerance challenges of spot preemption in serverless workflows unexplored.

\section{Conclusion}\label{sec:conclude}

We presented \sys, a serverless workflow execution system that addresses the cascade failures spot preemptions trigger in complex DAG workflows. 
Prior systems protect high-impact tasks by migrating them to reliable, non-spot resources, sacrificing spot's cost advantage. 
\sys instead classifies each task once at submission time and shares this classification across its three online mechanisms, protecting the same high-impact positions on spot, without runtime coordination or migration.
Our evaluation with \wfcnt DAG workflows on an in-house trace-driven spot emulator shows that \sys reaches an 89.03\% success rate at \$0.018 cost per success, outperforming four state-of-the-art baselines on success rate, cost, and latency simultaneously.

\begin{acks}
We thank the anonymous reviewers and our shepherd, Roy Friedman, for their feedback and guidance. We also thank Preetam Jain for his contributions to the early stage of this work. This work was supported in part by the National Science Foundation under grants 2416214 and 2443334.
\end{acks}

\bibliographystyle{ACM-Reference-Format}
\bibliography{bibs/reference.bib}

\end{document}